\documentclass{article}

\usepackage{graphicx}
\usepackage{xspace}
\usepackage{url}
\usepackage{amssymb}
\usepackage{latexsym}
\usepackage{amsthm}
\usepackage{amsmath}
\usepackage{amstext}
\usepackage{times}
\usepackage{calc}
\usepackage{amsopn}
\usepackage[noend]{algorithmic}
\usepackage[boxed]{algorithm}
\usepackage{eucal}
\usepackage{latexsym}
\usepackage{tabularx}
\usepackage{todonotes}
\usepackage{multirow}
\usepackage{tikz}

\floatname{algorithm}{Pseudocode}

\newcommand{\Oh}{\ensuremath{\mathcal{O}}}
\newcommand{\splitlong}{\textsc{Split Vertex Deletion}\xspace}
\newcommand{\splitvd}{\textsc{SplitVD}\xspace}
\newcommand{\vertexcover}{\textsc{Vertex Cover}\xspace}

\newtheorem{theorem}{Theorem}[section]

\newtheorem{corollary}[theorem]{Corollary}

\theoremstyle{definition}

\begin{document}

  \date{}

  \author{
  Marek Cygan
  \thanks{
    IDSIA, University of Lugano, Switzerland, \texttt{marek@idsia.ch}.
    Partially supported by the ERC Starting Grant NEWNET 279352 and Foundation for Polish Science.
  }
  \and
  Marcin Pilipczuk 
  \thanks{
    Institute of Informatics, University of Warsaw, Poland, \texttt{malcin@mimuw.edu.pl},
    Partially supported by NCN grant N206567140 and Foundation for Polish Science.
  }
  }

  \title{On fixed-parameter algorithms for \splitlong}

\maketitle

\begin{abstract}
In the \splitlong problem, given a graph $G$ and an integer $k$,
we ask whether one can delete $k$ vertices from the graph $G$ to obtain
a {\em{split graph}} (i.e., a graph, whose vertex set can be partitioned into two sets:
    one inducing a clique and the second one inducing an independent set).
In this paper we study fixed-parameter algorithms for \splitlong parameterized by $k$:
we show that, up to a factor quasipolynomial in $k$ and polynomial in $n$, the \splitlong
problem can be solved in the same time as the well-studied \vertexcover problem.
Plugging the currently best fixed-parameter algorithm for \vertexcover due to Chen et al. [TCS 2010],
we obtain an algorithm that solves \splitlong in time $\Oh(1.2738^k k^{\Oh(\log k)} + n^{\Oh(1)})$.

To achieve our goal, we prove the following structural result that may be of independent interest:
for any graph $G$ we may compute a family $\mathcal{P}$ of size $n^{\Oh(\log n)}$ 
containing partitions of $V(G)$ into two parts, such
for any two disjoint sets $X_C, X_I \subseteq V(G)$ where $G[X_C]$ is a clique
and $G[X_I]$ is an independent set, there is a partition in $\mathcal{P}$ which contains
all vertices of $X_C$ on one side and all vertices of $X_I$ on the other.
\end{abstract}

\section{Introduction}

The family of vertex deletion, or, more generally,
graph modification problems, has been studied very intensively,
both in theory and in practice.
As in many cases we expect the number of allowed modifications to
be small, compared to the size of the input graph,
   and most graph modification problems
turned out to be NP-hard (e.g., all vertex deletion problems for nontrivial hereditary
graph classes, by the classical
result of Lewis and Yannakakis \cite{lewis-yannakakis}),
it is natural to study these problems from the parameterized point of view, considering
parameterization by the solution size (the number of allowed modifications).

In the parameterized setting we assume that each instance is equipped with 
an additional value $k$ --- a parameter which aims to reflect
the instance complexity.
The goal is to provide an algorithm (called a fixed-parameter algorithm) with $f(k)n^{\Oh(1)}$
time complexity, where $n$ is the instance size and $f$
is a function independent of $n$.
Observe that such an algorithm is polynomial for any constant value of $k$
and moreover the degree of the polynomial is independent of the parameter value.
For more information about the parameterized complexity in general,
we refer to three monographs~\cite{downey-fellows:book, grohe:book, niedermeier:book}.

In this paper we focus on one particular graph modification problem, namely
the \splitlong problem (\splitvd for short). Here, we are given an $n$-vertex graph
$G$ and an integer $k$ and the task is to delete $k$ vertices from $G$
to obtain a {\em{split graph}}: a graph $H$ is called a {\em{split graph}}
if $V(H)$ can be partitioned into two parts $X_C$ and $X_I$, such that $H[X_C]$
is a clique and $H[X_I]$ is an independent set.\footnote{Through the paper
  we use standard graph notation, see e.g. \cite{diestel}.
  In particular, for a given graph $G$, by $V(G)$ and $E(G)$ we denote
  its vertex and edge set, respectively. For a set $X \subseteq V(G)$,
  $G[X]$ is a subgraph induced by $X$. For a vertex $v \in V(G)$, $N_G(v)$ denotes the set of
  neighbours of $v$ and $N_G[v] = N_G(v) \cup \{v\}$.}
Note that the partition $(X_C,X_I)$ does not need to be unique; for example,
an $n$-vertex clique is a split graph with $n+1$ different valid partitions.

As the class of split graphs is hereditary, by the result of Lewis and Yannakakis \cite{lewis-yannakakis}, \splitvd is NP-hard.
F\"{o}ldes and Hammer \cite{hammer} proved that the class of split graphs
is exactly the class of $\{2K_2,C_4,C_5\}$-free graphs; by the general
result of Cai \cite{cai:vd}, this observation yields a fixed-parameter
algorithm with running time $\Oh(5^k n^{\Oh(1)})$.
The dependency on $k$ has been subsequently improved
to $\Oh(2.32^k n^{\Oh(1)})$ by Lokshtanov et al. \cite{saket:lp}
and $\Oh(2^k n^{\Oh(1)})$ by Ghosh et al. \cite{ashutosh}.
In this paper we show that \splitvd can be solved essentially
in the same time as the well-studied \vertexcover problem.

\begin{theorem}\label{thm:alg}
If there exists an algorithm that solves the \vertexcover problem parameterized
by the solution size $k$ on $n$-vertex graphs in $f(k,n)$ time and $g(k,n)$ space,
then the \splitlong problem on $n$-vertex graphs can be solved
in $\Oh(f(k,n) k^{\Oh(\log k)} + n^{\Oh(1)})$ time and $\Oh(g(k,n) + n^{\Oh(1)})$ space.
\end{theorem}

By plugging in the currently fastest known algorithm for \vertexcover \cite{vc:best},
   we obtain the following.

\begin{corollary}
The \splitlong problem can be solved in 
\linebreak 
$\Oh(1.2738^k k^{\Oh(\log k)} + n^{\Oh(1)})$ time and polynomial space.
\end{corollary}

Note that there exists a straightforward reverse reduction:
given a \vertexcover instance $(G,k)$ (i.e., we ask for a vertex cover of size $k$
in the graph $G$), it is easy to see that an equivalent
\splitvd instance $(G',k)$ can be created by defining the graph $G'$ to be
a disjoint union of the graph $G$ and a clique on $k+2$ vertices.
Thus, we obtain that --- up to a factor quasipolynomial in $k$ and polynomial in $n$ ---
the optimal
time complexities of fixed-parameter algorithms for \vertexcover and \splitvd are equal.

The core difficulty of the proof of Theorem \ref{thm:alg} lies in the following
structural result that may be of independent interest.
\begin{theorem}\label{thm:parts}
For any $n$-vertex graph $G$ there exists a family $\mathcal{P}$
of partitions $(V_C,V_I)$ of the vertex set $V(G)$, such that
the following holds.
\begin{enumerate}
\item For any set $X \subseteq V(G)$ such that $G[X]$ is a split graph,
  and any partition $(X_C,X_I)$ of $X$, such that $G[X_C]$ is a clique
  and $G[X_I]$ is an independent set, there exists
  a partition $(V_C,V_I) \in \mathcal{P}$ such that
  $X_C \subseteq V_C$ and $X_I \subseteq V_I$.
\item $|\mathcal{P}| \leq 4 \cdot (2n)^{2\lfloor \log n \rfloor + 1}$.
\end{enumerate}
Moreover, there exists an algorithm that enumerates (with possible repetitions) the family $\mathcal{P}$
and runs in time $\Oh(n^{2 \lfloor \log n \rfloor + \Oh(1)})$ and polynomial space.
\end{theorem}

Theorem \ref{thm:parts} is proven in Section \ref{sec:parts}.
Equipped with this structural result, in Section~\ref{sec:proof-1.1} we
show that Theorem \ref{thm:alg} follows
easily by combining an already known preprocessing routine for \splitvd that
outputs an equivalent instance of size polynomial in $k$ (called a {\em{polynomial kernel}}),
Theorem \ref{thm:parts} and a simple observation that, if we seek for a resulting
split induced subgraph that is covered by a fixed
partition $(V_C,V_I) \in \mathcal{P}$, \splitvd naturally reduces to a \vertexcover
instance with the same parameter.

\section{Small family of reasonable partitions: proof of Theorem \ref{thm:parts}}\label{sec:parts}
\newcommand{\stan}{\ensuremath{\mathcal{S}}}

In this section we prove Theorem \ref{thm:parts}.
To this end, we describe a branching algorithm that
computes the family $\mathcal{P}$. The algorithm
maintains a partition (called a {\em{state}}) of $V(G)$ into three parts $V_C^0$, $V_I^0$ and $A$;
intuitively, the vertices of $V_C^0$ and $V_I^0$ are already assigned
to $V_C$ and $V_I$, whereas the set $A$ consists of remaining ({\em{active}}) vertices.
At each step, given a state $\stan = (V_C^0,V_I^0,A)$, the algorithm
outputs two partitions $(V_C^0 \cup A, V_I^0)$ and $(V_C^0,V_I^0 \cup A)$
and branches (calls itself recursively) into $2|A|$ subcases,
creating two new states for each $v \in A$:
a state
$\stan_{v \to C} = (V_C^0 \cup \{v\}, V_I^0 \cup (A \setminus N_G[v]), A \cap N_G(v))$
and a state
$\stan_{v \to I} = (V_C^0 \cup (A \cap N_G(v)), V_I^0 \cup \{v\}, A \setminus N_G[v])$.
Informally speaking, in the first branch the vertex $v$ is assigned to the clique part; consequently,
all its non-neighbours are assigned to the independent set part, as they cannot be together with $v$
in the clique part of a split induced subgraph of $G$. The second branch symmetrically assigns
$v$ to the independent set part and all neighbours of $v$ to the clique part.

Moreover, the recurrence is trimmed at depth $2\lfloor \log n \rfloor + 1$.
The algorithm is described on Pseudocode \ref{alg:parts}.

\begin{algorithm}
\begin{algorithmic}[1]
\REQUIRE $\mathrm{Generator}(G,d,\stan=(V_C^0,V_I^0,A))$
\COMMENT{$n=|V(G)|$ and $\stan=(V_C^0,V_I^0,A)$ is a partition of $V(G)$}
\STATE output $(V_C^0 \cup A, V_I^0)$ and $(V_C^0,V_I^0 \cup A)$.
\IF{$d < 2 \lfloor \log n \rfloor + 1$}
  \FORALL{vertices $v \in A$}
    \STATE $\mathrm{Generator}(G,d+1,\stan_{v \to C} = (V_C^0 \cup \{v\}, V_I^0 \cup (A \setminus N_G[v]), A \cap N_G(v)))$
    \STATE $\mathrm{Generator}(G,d+1,\stan_{v \to I} = (V_C^0 \cup (A \cap N_G(v)), V_I^0 \cup \{v\}, A \setminus N_G[v]))$
  \ENDFOR
\ENDIF

\REQUIRE $\mathrm{GeneratePartitions}(G)$
\STATE $\mathrm{Generator}(G,0,(\emptyset,\emptyset,V(G)))$.
\end{algorithmic}
\caption{Algorithm that generates the family $\mathcal{P}$ from Theorem \ref{thm:parts}.}\label{alg:parts}
\end{algorithm}

Since the algorithm trims the recurrence at depth $2 \lfloor \log n \rfloor + 1$, the bounds on the running time
and the size of the family $\mathcal{P}$ follow: at each step, $2|A| \leq 2n$ new subcases are created,
the search tree contains at most $(2n)^{2\lfloor \log n \rfloor + 1}$ leaves and less than twice as much vertices,
and each call to the procedure $\mathrm{Generator}$ outputs two partitions.
To finish the proof of Theorem \ref{thm:parts}, we need to show the computed family $\mathcal{P}$
admits the first property of Theorem~\ref{thm:parts}.

To this end, let us fix a set $X \subseteq V(G)$ that induces a split graph in $G$ and a partition $(X_C,X_I)$
of $X$ such that $G[X_C]$ is a clique and $G[X_I]$ is an independent set. We show that the algorithm outputs
a partition $(V_C,V_I)$ with $X_C \subseteq V_C$ and $X_I \subseteq V_I$.

We say that a state $\stan = (V_C^0,V_I^0,A)$ is {\em{promising}} if $X_C \subseteq V_C^0 \cup A$ and $X_I \subseteq V_I^0 \cup A$;
note that this is a necessary condition to output a desired partition in any subcase generated from the state $\stan$.
Moreover, note that the initial state $(\emptyset,\emptyset,V(G))$ is clearly promising.

Consider a promising state $\stan = (V_C^0,V_I^0,A)$.
Denote $X_C^A = X_C \cap A$ and $X_I^A = X_I \cap A$. Note that if $X_C^A = \emptyset$, then the
partition $(V_C^0, V_I^0 \cup A)$ is a desired partition. Symmetrically, if $X_I^A = \emptyset$, then the partition
$(V_C^0 \cup A, V_I^0)$ is a desired partition; both these partitions are output by the algorithm.

Consider now the remaining case where $X_C^A$ and $X_I^A$ are nonempty.
Note that for $v \in X_C^A$, the state
$\stan_{v \to C}$ is also promising, as $G[X_C]$ is a clique and $X_C \subseteq \{v\} \cup N_G(v)$.
Symmetrically, for any $v \in X_I^A$, the state $\stan_{v \to I}$ is also promising,
as $G[X_I]$ is an independent set and $X_I \subseteq V(G) \setminus N_G(v)$.
However, our recurrence is trimmed at depth $2 \lfloor \log n \rfloor + 1$.
To cope with this obstacle, we show that
there exists a choice of $v \in A$ that efficiently reduces the sizes of $X_C^A$ and $X_I^A$.

Let $F$ be the set of edges of $G$
that have one endpoint in $X_C^A$ and second endpoint in $X_I^A$. If $|F| > |X_C^A| \cdot |X_I^A| / 2$
(i.e., there are more edges between $X_C^A$ and $X_I^A$ than non-edges) then,
by standard averaging argument, there exists a vertex $v \in X_I^A$ such that $|N_G(v) \cap X_C^A| > |X_C^A|/2$ (i.e., more than
 half of the vertices of $X_C^A$ are neighbours of $v$)
Otherwise, if $|F| \leq |X_C^A| \cdot |X_I^A| / 2$, then there exists a vertex $v \in X_C^A$ such that $|X_I^A \setminus N_G(v)| \geq |X_I^A|/2$
(i.e., at least half of the vertices of $X_I^A$ are not neighbours of $v$).
In the first case, in the promising state $\stan_{v \to I}$ the size of the set $X_C^A$ is reduced by at least half; in the second case,
in the promising state $\stan_{v \to C}$ the size of the set $X_I^A$ is reduced by at least half. 
At the beginning, $|X_C^A|,|X_I^A| \leq n$, thus the recurrence reaches a promising state where $X_C^A$ or $X_I^A$ is empty at depth
at most $2 \lfloor \log n \rfloor +1$. This finishes the proof of Theorem \ref{thm:parts}.

\section{The algorithm: proof of Theorem \ref{thm:alg}}\label{sec:alg}
\label{sec:proof-1.1}

Equipped with Theorem \ref{thm:parts}, we are now ready to show the proof of Theorem \ref{thm:alg}.
  Consider a \splitvd instance~$(G,k)$.
First, we invoke one of the known preprocessing (kernelization) routines for \splitvd that reduces the number of vertices of the graph to a polynomial in $k$, without increasing the parameter.
Here, we can either use the generic framework of the $d$-\textsc{Hitting Set} problem \cite{hitting-set}
(recall that the class of split graphs has a finite set of forbidden induced subgraphs)
or use the recent $\Oh(k^3)$-vertex kernel by Ghosh et al. \cite{ashutosh}.
This step adds an additive factor of polynomial order in $|V(G)|$ both to time and space complexity of the algorithm.

Second, we invoke Theorem \ref{thm:parts} and process the output partitions one by one.
For a given partition $(V_C,V_I)$, we seek for a set $X \subseteq V(G)$, such that $G[V_C \cap X]$ is a clique,
$G[V_I \cap X]$ is an independent set and $|V(G) \setminus X| \leq k$.
By Theorem \ref{thm:parts} this is sufficient to solve the initial \splitvd instance $(G,k)$,
and this step adds an $k^{\Oh(\log k)}$ multiplicative factor to the time complexity and
a polynomial in $k$ additive factor to the space complexity.

Fix a partition $(V_C,V_I)$. We are to delete at most $k$ vertices from the graph $G$
to make $G[V_C]$ a clique and $G[V_I]$ an independent set.
Let $G'$ be defined as a disjoint union of $G[V_I]$ and a complement of $G[V_C]$.
Note that our task becomes the classical vertex cover problem in the graph $G'$ with parameter $k$: we need to
cover all edges of $G[V_I]$ and non-edges of $G[V_C]$. Therefore, for a fixed partition $(V_C,V_I)$,
the problem can be solved in the same time as the \vertexcover problem for a graph of the same size and parameter $k$.
This finishes the proof of Theorem \ref{thm:alg}.

\section{Conclusions}

We have shown that the dependencies on the parameter $k$ in the optimal time complexity of fixed-parameter algorithms
for \vertexcover and \splitlong are essentially equal.
This result can be considered as a tight bound on the time complexity of fixed-parameter algorithms for \splitlong.

However, note that our reduction adds a polynomial in the size of the input graph additive factor to the time complexity
that results from the application of a kernelization algorithm.
The algorithm of Chen et al. \cite{vc:best} for the \vertexcover problem has linear dependency on $n$.
We leave as an open problem to obtain a linear-time polynomial kernel for \splitvd; such a result would
automatically yield a linear-time dependency on $n$ in our algorithm.

\bibliographystyle{plain}
\bibliography{split-vd}

\end{document}